\documentclass[final,3p,times,twocolumn]{elsarticle}
\usepackage{graphics}
\usepackage{graphicx}
\usepackage{epsfig}
\usepackage{amssymb}
\usepackage{amsthm}
\usepackage{lineno}
\usepackage{amsmath}

\newcommand{\Xp}{\Xi_{cc}^+}
\newcommand{\Xpp}{\Xi_{cc}^{++}}
\newcommand{\Lc}{\Lambda_c^+}
\newcommand{\Jp}{J/\psi}
\newcommand{\GeV}{\mbox{\,GeV}}
\newcommand{\fb}{\mbox{\,fb}}

\begin{document}

\begin{frontmatter}

\title{Production of doubly charmed baryons nearly at rest}

\author{Stefan Groote, Sergey Koshkarev}
\address{Institute of Physics, University of Tartu, Tartu 51010, Estonia}

\begin{abstract}
We investigate the production cross sections, momentum distributions and
rapidity distributions for doubly charmed baryons which according to the
intrinsic heavy quark mechanism are produced nearly at rest. These events
should be measurable at fixed-target experiments like STAR@RHIC and AFTER@LHC.
\end{abstract}


\end{frontmatter}

\section{Introduction\label{intro}}

Doubly heavy baryons are a rigorous prediction of quantum chromodynamics. In
the 2000's, evidence for the existence of the doubly charmed baryons was
reported by the SELEX Collaboration~\cite{Mattson2002,MattsonPhD,%
Moinester2003,Ocherashvili2005,Engelfried2005,Engelfried2006}. However, the
discrepancy of the production properties derived from the SELEX data with
predictions of perturbative QCD induced wrong expectations for the production
rates of the doubly charmed baryons in the different production environments,
leading to reports stating the non-evidence of doubly heavy
baryons~\cite{FOCUS,BaBar,Belle,LHCb}.

In Ref.~\cite{Anikeev2017} it is shown that the intrinsic charm mechanism can
fill the gap between the rates motivated by the SELEX data and the
theoretically predicted production rates. Moreover, the ratio
$\sigma( c \bar c c \bar c)/\sigma(c \bar c)$ calculated from the SELEX data
is comparable or even smaller than the same rate derived from the double $\Jp$
production data at the NA3 experiment~\cite{Badier1983,Badier1985}. In
addition, the intrinsic charm mechanism can perfectly explain the kinematics
of the doubly charmed baryons with $\langle x_F \rangle \sim 0.33$ and the
relatively small mean transverse momentum of approx.\ $1\GeV/c$.

The most recent Belle and LHCb results cannot argue against the SELEX data.
The Belle experiment~\cite{Belle} presented an upper limit on the cross
section $\sigma(e^+ e^- \to \Xp + X)$ of $82-500\fb$ at 95\% confidence level
for the decay mode with $\Lc$ at $\sqrt{s} = 10.58\GeV$, using a luminosity of
$980\fb^{-1}$. This turns out to be at least twice as much as the theoretical
upper limit given by $\sigma(\Xp)\approx 35\pm 10\fb$~\cite{UFN,Kiselev1994}.

The LHCb Collaboration~\cite{LHCb} published upper limits at 95\% confidence
level on the ratio $\sigma(\Xp)\cdot Br(\Xp\to\Lc K^-\pi^+)/\sigma(\Lc)$ of
$1.5\times 10^{-2}$ and $3.9\times 10^{-4}$ for the lifetimes of 100~fs and
400~fs, respectively, and for an integrated luminosity of $0.65\fb^{-1}$. This
is comparable with results from Refs.~\cite{UFN,ChangPRD,Chang2006,Gunter} of
about $10^{-4}-10^{-3}$. However, the minimum lifetime reached by the LHCb is
about three times lager than the one measured by the SELEX experiment which is
$\tau(\Xp)<33$~fs at 90\% confidence level~\cite{Mattson2002}, and almost two
times larger than the theoretical prediction of
$\tau(\Xp)\approx 53$~fs~\cite{Karliner2014}. In other words, the LHCb
Collaboration provided an analysis outside the signal region.

The production of doubly heavy baryons at high Feynman-$x$ at the scheduled
future fixed-target experiment at the LHC (AFTER@LHC) via the intrinsic heavy
quark mechanism is already discussed in Refs.~\cite{Anikeev2017,Koshkarev2017}.
In addition, some existing experiments have fixed-target
programs~\cite{STAR2016,SMOG2012,SMOG2015}. However, as these experiments were
built as collider detectors, they have very limited access to high Feynman-$x$.

In a recent talk~\cite{Brodsky2017}, Stanley Brodsky proposed the production
of charmed hadrons via the intrinsic charm mechanism from the target (see also
chapter~6.1 in Ref~\cite{Brodsky2016}). In this paper we investigate the
opportunity to produce doubly charmed baryons via the intrinsic heavy quark
mechanism from the target at the fixed-target experiment at
STAR~\cite{STAR2016}, the fixed-target detector at the LHCb experiment
(SMOG@LHCb)~\cite{SMOG2012,SMOG2015} and at the scheduled future fixed-target
experiment at the LHC (AFTER@LHC)~\cite{Brodsky2013,Lansberg2012}. 

\section{Revisiting the SELEX data\label{selex}}

The SELEX experiment was a fixed-target experiment utilizing the Fermilab
negative and positive charged beams at $600\GeV/c$ to produce charm particles
in a set of thin foil of Cu or in a diamond. It was operated in the kinematic
region $x_F > 0.1$. The negative beam composition was about 50\% $\Sigma^-$
and 50\% $\pi^-$ while the positive beam was composed of 90\% protons. The
experimental data recorded used both positive and negative beams. 67\% of the
events were induced by $\Sigma^-$, 13\% by $\pi^-$, and 18\% by protons.

The production cross section was not provided by the SELEX collaboration.
However, the production properties of the doubly charmed baryons can be
compared to that of the $\Lc$ baryon. To simplify the analysis we will take a
look only at the production of the 20 signal events for the decay mode
$\Xpp \to \Lc K^- \pi^+ \pi^+$ at a mass of $3.76\GeV$ over a sample of 1656
events for $\Lc\to pK^-\pi^+$~\cite{MattsonPhD} with $x_F(\Lc)>0.15$. This
sample was previously used for a precision measurement of the lifetime of
$\Lc$~\cite{SELEX2000,Kushnirenko}. The measured production ratio $R$ is
defined by
\[
R = \frac{\sigma(\Xpp) \cdot Br(\Xpp \to \Lc K^- \pi^+ \pi^+)}{\sigma(\Lc)}
=\frac{N_{\Xpp}}{\epsilon_{++}} \cdot \frac{\epsilon_{\Lc}}{N_{\Lc}},
\]
\begin{figure}[t]
\includegraphics[scale=0.35] {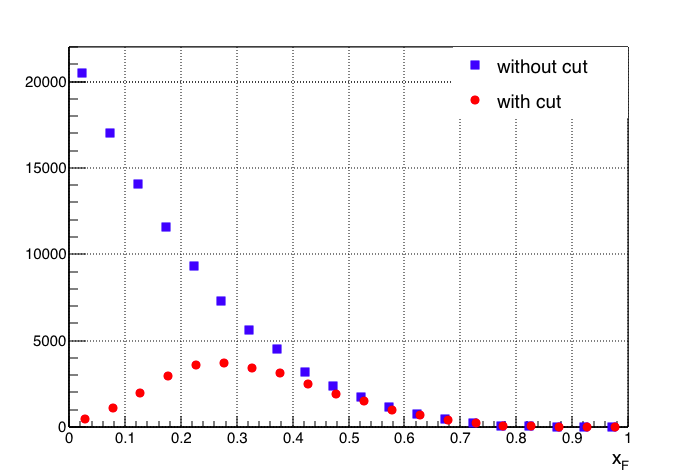}
\caption{\label{fig:pQCDaccept}The squared (blue) points represent the pQCD
motivated $x_F$ distribution of $\Xpp$ baryons. The circular (red) points show
this distribution with the experimental geometry cut (cf.\
Ref.~\cite{MattsonPhD}).}
\end{figure}
\begin{figure}[t]
\includegraphics[scale=0.35] {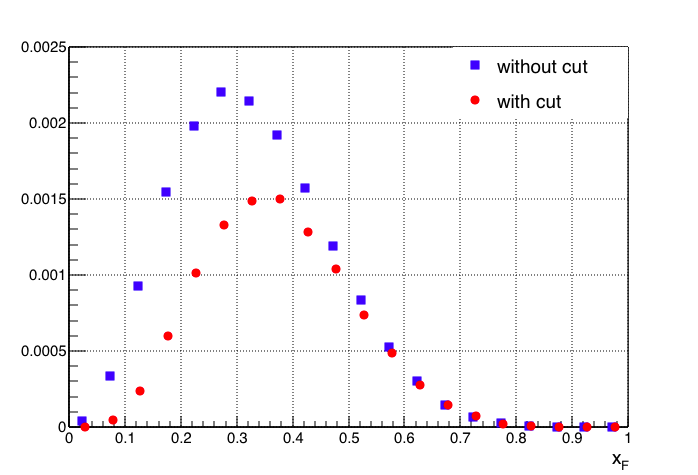}
\caption{\label{fig:ICaccept}The squared points represent the IC motivated
$x_F$ distribution of $\Xpp$ baryons. The circular point show this
distribution re-weighted with the experimental reconstruction efficiency as
function of $x_F$. The reconstruction efficiency is extracted from
Fig.~\ref{fig:pQCDaccept}.}
\end{figure}
where $N$ is the number of events in the respective sample and the
reconstruction efficiency of $\Xpp$ is given by
$1/\epsilon_{++} \simeq 3.7$~\cite{MattsonPhD}. Central values for the number
$N_{\Lc}/\epsilon_{\Lc}$ of corrected events can be found in Ref.~\cite{Garcia}
to lie between $13326$ and $10010$ according whether the lowest bin with
$x_F\in[0.125,0.175]$ is taken into account or not. However, using the
intrinsic charm mechanism as the production mechanism, $\epsilon_{++}$ will be
at least $2.3$ times bigger (see Figs.~\ref{fig:pQCDaccept}
and~\ref{fig:ICaccept} for the acceptance for perturbative QCD and intrinsic
charm, respectively). Therefore, we obtain
\begin{equation*}
R \sim (2-3) \times 10^{-3}.
\end{equation*}
This value is at least an order of magnitude smaller than the production ratio
in the sample $N_{\Xpp}/\epsilon_{++}\cdot 1/N_{\Lc}\approx 0.045$ which is
usually erroneously used as the production ratio.

\section{Theoretical background\label{base}}

Origin, properties and possibilities for the detection of a non-perturbative
intrinsic heavy flavor component in the nucleon are widely discussed in the
literature~\cite{Brodsky1984,Franz,Brodsky1980,Brodsky1981,Vogt1995,%
Brodsky2013,Lansberg2012,Rakotozafindrabe,Brodsky2015}. QCD predicts such
components from the outset. Intrinsic charm and bottom quarks are contained in
the wavefunction of a light hadron and originate from diagrams where the heavy
quarks are multiply attached via gluons to the valence quarks. Intrinsic heavy
flavor components are contributed by the twist-six contribution of the
operator product expansion proportional to $1/m_Q^2$~\cite{Brodsky1984,Franz}.
In this case, the frame-independent light-front wavefunction of the light
hadron has maximum probability if the Fock state is minimally off-shell. This
means that all the constituents are at rest in the hadron rest frame and thus
have the same rapidity $y$ if the hadron is boosted. Equal rapidity occurs if
the light-front momentum fractions $x=k^+/P^+$ of the Fock state constituents
are proportional to their transverse masses,
$x_i\propto m_{T,i}=(m^2_i+k^2_{T,i})^{1/2}$, i.e.\ if the heavy constituents
have the largest momentum fractions. This features the BHPS model given by
Brodsky, Hoyer, Peterson and Sakai for the distribution of intrinsic heavy
quarks~\cite{Brodsky1980,Brodsky1981}.

In the BHPS model the wavefunction of a hadron in QCD can be represented as a
superposition of Fock state fluctuations, e.g.\ $| h \rangle \sim | h_l
\rangle + | h_l g \rangle + | h_l Q \bar{Q} \rangle \ldots$, where $ h_l$ is
the light quark content, and $Q=c,b$. If the projectile interacts with the
target, the coherence of the Fock components is broken and the fluctuation can
hadronize. The intrinsic heavy flavor Fock components are generated by virtual
interactions such as $gg \to Q \bar{Q}$ where the gluons couple to two or more
valence quarks of the projectile. The probability to produce such $ Q \bar{Q}$
fluctuations scales as $\alpha_s^2 (m_Q^2)/m_Q^2$ relative to the
leading-twist production.

Following Refs.~\cite{Brodsky1980,Brodsky1981,Vogt1995}, the general formula
for the probability distribution of two heavy quark pairs by intrinsic heavy
flavor Fock state as a function of the momentum fractions $x_i$ is given by
\begin{equation}\label{eq:IC}
\begin{aligned}
 \frac{dP_{iQ_1Q_2}}{\prod_{i=1}^n dx_i} \propto
  \alpha_s^4 (M_{Q_1 \bar{Q}_1}) \alpha_s^4 (M_{Q_2 \bar{Q}_2})
  \frac{\delta \big( 1 - \sum_{i=1}^n x_i \big)}{\big(\sum_{i=1}^n
  \hat{m}_{T,i}^2 / x_i \big)^2},
\end{aligned}
\end{equation}
where $\hat{m}_i=(m_i^2+\langle k^2_{T,i}\rangle)^{1/2}$ is the effective
mass, $k^2_{T,i}$ is the mean transverse momenta, and the masses of the light
quarks are neglected.

\begin{figure}[t]
\includegraphics[scale=0.35] {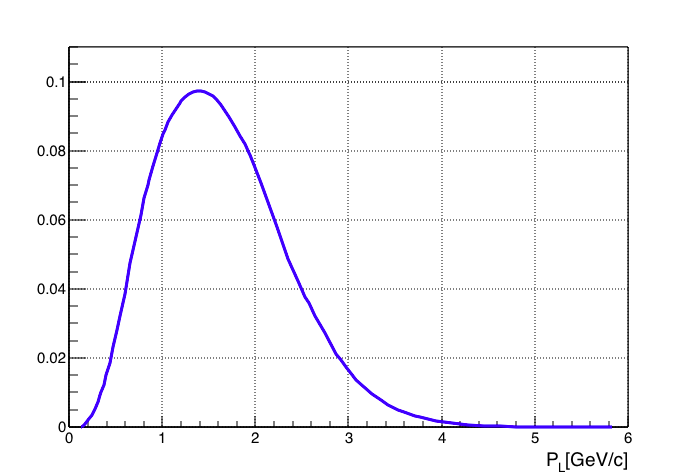}
\caption{\label{fig:PL}Momentum distribution of the $\Xi_{cc}$ baryons
produced by the intrinsic charm from the target with a $200\GeV/c$ proton
beam}
\end{figure}
\begin{figure}[t]
\includegraphics[scale=0.35] {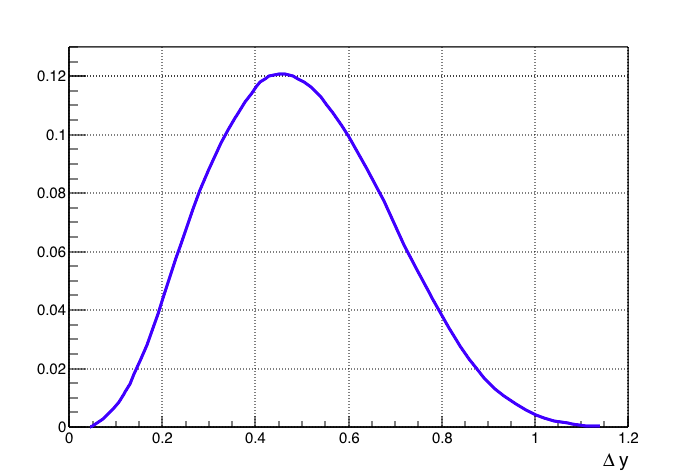}
\caption{\label{fig:rapid}Rapidity difference of the $\Xi_{cc}$ baryons
produced by the intrinsic charm from the target with a $200\GeV/c$ proton
beam}
\end{figure}

The normalization of the production cross section of two charm pairs is given
by
\begin{equation}\label{eq:bbb}
\sigma_{icc} = \frac{P_{icc}}{P_{ic}} \cdot \sigma_{ic},\qquad
\sigma_{ic} = P_{ic} \cdot  \sigma^{\it inel} \frac{\mu^2}{4 \hat{m}_c^2}\,,
\end{equation}
where $\mu^2\approx 0.2\GeV^2$ denotes the squared soft interaction scale
parameter, the effective transverse $c$-quark mass is given by
$\hat{m}_c=1.5\GeV$, and $\sigma^{\it inel}$ is the inelastic proton--proton
cross section.

The nuclear dependence scaling from the manifestation of intrinsic charm is
expected to be $\sigma_A\approx\sigma_{icc}\cdot A^{2/3}$ for production from
the beam and $\sigma_A\approx\sigma_{icc}\cdot A$ for production from the
target~\cite{Brodsky2017}.

\section{Production of the doubly charmed baryons\label{crsec}}

The production cross section of the doubly charmed baryon can be obtained as
an application of the principle of quark--hadron duality. According to this
principle, the cross section of the baryon is obtained by calculating the
production of a $QQ$ pair in the small invariant mass interval between $2m_Q$
and the threshold to produce open heavy quark hadrons, $2m_H$. The $QQ$ pair
has $3\times3=(\bar3+6)$ color components, consisting of a color antitriplet
and a color sextet. The probability that a $QQ$ pair forms an antitriplet
state is $\bar3/(\bar3+6)=1/3$. Therefore, in case of the doubly charmed
baryon the cross section will be
\begin{equation}\label{eq:ccc}
\sigma(cc) = \frac{1}{3} f_{cc} \sigma_{icc},
\end{equation}
where $f_{cc}$ is the fragmentation ratio of the $cc$ pair written as
\begin{equation}\label{eq:ddd}
f^{icc}_{cc} = \int_{4m_c^2}^{4m_D^2} dM_{cc}^2
  \frac{dP_{icc}}{dM_{cc}^2} \,\, \bigg/
  \int_{4m_c^2}^{s} dM_{cc}^2 \frac{dP_{icc}}{dM_{cc}^2} .
\end{equation}

Not all $cc$ pairs form a doubly charmed baryon. Unfortunately, the
fragmentation rate is unknown. However, if we consider the diquark $cc$ as
heavy antiquark and analyze the fragmentation of $c$ and $b$ quarks into
mesons, for example $f(c \to D^+)=0.217\pm 0.043$~\cite{ZEUS2005} or
$f(b\to B^+,B^0)=0.344\pm 0.021$~\cite{PDG2017}, we can assume the doubly
charmed baryonic cross section to be
\begin{equation}
\sigma(\Xi_{cc}) \sim (0.2 - 0.3) \cdot \sigma(cc).
\end{equation}

The STAR fixed-target program is a fixed-target experiment using the proton
beam of the Relativistic Heavy Ion Collider (RHIC) up to $250\GeV/c$ and the
Au beam up to $100\GeV/c$ colliding with a wired target. Combining
Eqs.~(\ref{eq:bbb}), (\ref{eq:ccc}) and~(\ref{eq:ddd}) and using
$\sigma^{\it inel}_{pp}(p_{\rm beam}=200\GeV)\approx
32\text{\,mb}$~\cite{Schiz1981}, we may expect the production cross section of
the $\Xi_{cc}$ to be
\begin{equation*}
\sigma(\Xi_{cc})\approx(0.2-0.3)\times 75\,\text{nb}.
\end{equation*}

The kinematic limits on the energy and the momentum of the doubly charmed
baryon formed by the intrinsic charm from the target are given by
\begin{equation}\label{eq:kin}
E_{\it lab} = \frac{1}{2m_{\it tar}} (m_{cc}^2 + m_{\it tar}^2),\quad
p_{\it lab} = \frac{1}{2 m_{\it tar}} (m_{cc}^2 - m_{\it tar}^2).
\end{equation}
These last expressions depend solely on the two masses $m_{cc}$ and
$m_{\it tar}$ and no longer on the beam energy. Upon combining
Eqs.~(\ref{eq:IC}) and~(\ref{eq:kin}) we can find the momentum distribution
(Fig.~\ref{fig:PL}) and the distribution of the rapidity difference
$\Delta y=y-y_{\it tar}$ (Fig.~\ref{fig:rapid}) in the laboratory frame.
It is obvious that experiments at the STAR detector, typical at rapidities
$|y| < 1$ for track selection, have the potential to observe doubly charmed
baryons.

The SMOG@LHCb is a fixed-target experiment using the LHC beam at
$6500\GeV/c$ dumped in the Helium gas target. Following the logic from above
and using $\sigma^{\it inel}_{pp}(\sqrt{s} =110\,\text{GeV})\approx
37\mbox{\,mb}$~\cite{Block} we obtain the production cross section of the
$\Xi_{cc}$ to be
\begin{equation*}
\sigma(\Xi_{cc})\approx(0.2-0.3)\times 65\mbox{\,nb}.
\end{equation*}
The kinematic distributions have the same shape with slightly different mean
values. Unfortunately, SMOG@LHCb has acceptance only for $\Delta y > 2$ which
will make the detection of the doubly charmed baryons problematic.

AFTER@LHC is the scheduled future fixed-target experiment at the LHC operating
at $\sqrt{s} = 115\GeV$. Therefore, the estimation provided for SMOG@LHCb can
also be used for AFTER@LHC.

\section{Conclusion\label{conclusion}}

In this paper we investigated the beautiful prediction of the intrinsic heavy
quark mechanism. The doubly charmed baryons are produced from the target with
an approximate mean value for the momentum of about $1.5\GeV/c$. Such ``soft''
final states can be observed at the current and future fixed-target
experiments. The production cross sections are presented.

From the calculation we see that the intrinsic heavy quark mechanism does not
contribute to the region of negative $x_F$. The double intrinsic heavy quark
mechanism is not the leading production mechanism at the modern
accelerators~\cite{Groote2017}. However, it still can aim to searching exotic
states like double charmed baryons. In addition, we reinterpreted the SELEX
data and obtained a realistic production ratio for doubly charmed baryons
$(ccu)$ over $\Lc$ in the kinematic region of SELEX (see the discussion in
Sec.~\ref{selex}).

In the end it is interesting to note that using the fragmentation ratios of
$c$ and $b$ quarks into baryons, $f(c\to\Lc)=0.071\pm 0.21$~\cite{BaBar2007}
and $f(b\to b\text{-baryons})=0.197\pm 0.046$~\cite{PDG2017} we can also
roughly calculate the doubly, hidden and open charm, tetraquark production
cross section as $0.1\cdot\sigma(cc)$.

\subsection*{Acknowledgments}

We would like to thank S.J.~Brodsky for very detailed and productive
discussions on the production of heavy quark states from the target via the
intrinsic heavy quark mechanism. In addition, we would like to thank Guannan
Xie for comments on acceptance of the STAR detector and Yu.~Shcheglov for
comments on the acceptance of the LHCb detector. This work was supported by
the Estonian Research Council under Grant No.~IUT2-27.

\end{document}